\documentclass[fleqn,8pt]{wlscirep}
\newcommand{\ket}[1]{\vert #1 \rangle}
\newcommand{\bra}[1]{\langle #1 \vert}

\usepackage{amsmath}
\usepackage{amssymb}
\usepackage{graphicx}
\usepackage{dsfont}
\usepackage{epstopdf}
\usepackage{todonotes}
\usepackage{color}
\usepackage{ulem}
\usepackage{tocloft} 
\usepackage[fleqn]{nccmath} 
\usepackage{lmodern}
\usepackage{bm}
\usepackage{float}

\title{Nucleation of superfluid-light domains in a quenched dynamics}
\author[1,2]{Joaqu\'in Figueroa}
\author[1,2]{Jos\'e Rogan}
\author[1,2]{Juan Alejandro Valdivia}
\author[1,2]{Miguel Kiwi}
\author[3]{Guillermo Romero}
\author[1,2,$\dag$]{Felipe Torres}

\affil[1]{Departamento de F\'isica, Facultad de Ciencias,
  Universidad de Chile, Casilla 653, Santiago, Chile 7800024}
\affil[2]{Center for the Development of Nanoscience and Nanotechnology 9170124, Estaci\'on Central, Santiago, Chile}
\affil[3]{Departamento de F\'isica, Universidad de Santiago de Chile (USACH), 
	Avenida Ecuador 3493, 9170124, Santiago, Chile}
	
\affil[$\dag$]{felestorres@gmail.com}

\begin{abstract}
Strong correlation effects emerge from light-matter interactions in
  coupled resonator arrays, such as the Mott-insulator to superfluid
  phase transition of atom-photon excitations. We demonstrate that the
  quenched dynamics of a finite-sized complex array of coupled
  resonators induces a first-order like phase transition. The latter
  is accompanied by domain nucleation that can be used to manipulate
  the photonic transport properties of the simulated
  superfluid phase; this in turn leads to an empirical scaling law.  This universal
  behavior emerges from the light-matter interaction and the topology
  of the array. The validity of our results over a wide range of
  complex architectures might lead to a promising device for use in
  scaled quantum simulations.

\end{abstract}

\begin{document}

\flushbottom
\maketitle
%
%
\thispagestyle{empty}

\section*{Introduction}
The absence of energy dissipation in the flow dynamics of a quantum
fluid is one of the most fascinating effects of strongly correlated
condensates \cite{Nature135,QuantumLiquids_Book,
  Science269,PhysRevLett85,PhysRevLett91, PRLShiro}.  Quantum phase transitions,
from Mott insulator to superfluid, have been observed in a wide range
of physical platforms such as ultracold atoms in optical lattices
\cite{Nature415}, trapped gases of interacting fermionic atom pairs
\cite{PhysRevLett92}, and exciton-polariton condensates
\cite{NatPhys13, NatPhys6, NatPhys2014}.  Furthermore, the remarkable
progress in controlling light-matter interactions in the microwave
regime of circuit quantum electrodynamics (QED) has provided a
suitable scenario for studying strongly correlated effects with light
\cite{NatPhys8,PhysRevX4031043,PhysRevX7011016}.  In this case,
coupled resonator arrays (CRAs) each doped with a two-level system
(TLS) allow for the formation of dressed quantum states (polaritonic
states) and effective photon-photon interactions.  The underlying
physics is well described by the Jaynes-Cummings-Hubbard (JCH) model
\cite{Hartmann2006aa,PhysRevA76031805,NatPhys2006}.  In this case, if
the frequencies of the single resonator mode and the TLS are close to
resonance, the effective photonic repulsion prevents the presence of
more than one polaritonic excitation in the resonator, due to the
photon-blockade effect \cite{Nat7047, PhysRevLett79,JOpB}. Detuning
the atomic and photonic frequencies diminishes this effect and leads
the system to a photonic superfluid \cite{PhysRevA76031805}.  Unlike
Bose-Einstein condensation in optical lattices, polariton condensation
includes two kind of excitations, atomic and photonic, and the
transition from Mott-insulator to superfluid is accompanied by a
transition of the excitations from polaritonic to photonic
\cite{PhysRevA76031805}.

Here we show how a first-order like phase transition of the simulated
superfluid phase of polaritons in CRAs can be induced by a quench
dynamics as described by the JCH model.  We compare full numerical
simulations of several arrangements of CRAs with mean-field theory of
photonic fluctuations dynamics.  In this case, the simulated
Mott-superfluid transition relies on the topological properties of the
array, since the on-site photon blockade strongly depends on the
connectivity of each node, even for small resonator-resonator hopping
strength.  When the system is prepared in the Mott state with a
filling factor of one net excitation per site, and a sudden quench of
the detuning between the single resonator mode and the TLS is applied,
we find a first-order like phase transition which can be described by
two bosonic excitations of the lower and upper polariton band.  We
find that a nucleated superfluid photon state emerges in a localized
way, which depends on the topology of the array.  This avalanche-like
behavior near the simulated phase transition leads to a universal
scaling law between critical parameters of the superfluid phase and
the average connectivity of the array.

\bibliographystyle{apsrev}

\section*{The model} 
\begin{figure}[t]
\centering
\includegraphics[scale=0.35]{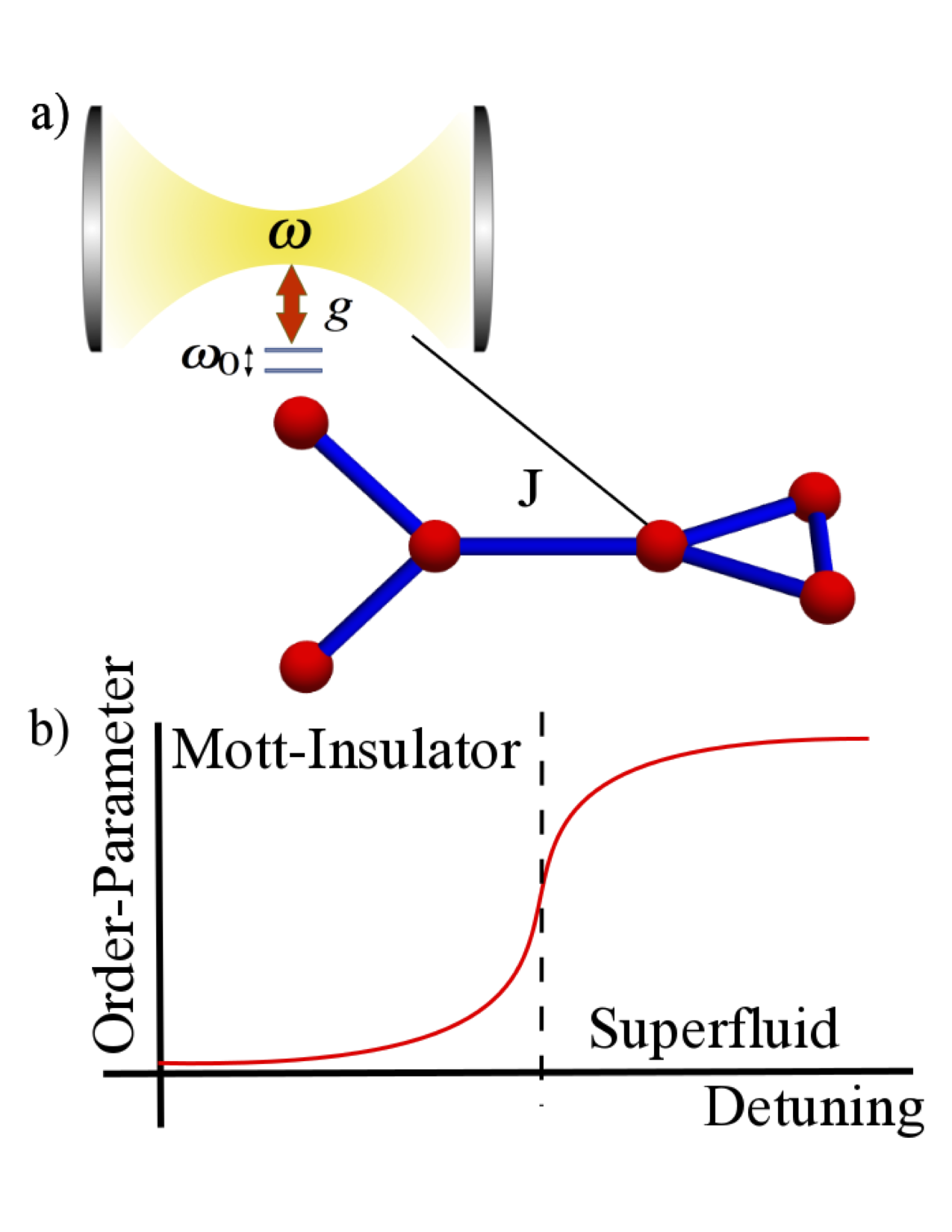}
\caption{(color online) a) Schematic representation of the
  Jaynes-Cummings-Hubbard lattice in a complex array where each node
  consists of a single resonator strongly coupled to a two-level
  system. b) Phase transition from Mott-insulator to superfluid in
  light-matter CRAs systems as a function of the detuning parameter.
}
\label{Fig1}
\end{figure}

The physical scenario that we consider are CRAs in complex
arrangements such as the one in Fig.~\ref{Fig1}(a). Here, each node of
the array consists of a QED resonator doped with a TLS to be a real or
artificial atom, and the whole system is described by the
Jaynes-Cummings-Hubbard model
\cite{Hartmann2006aa,PhysRevA76031805, NatPhys2006}, whose
Hamiltonian reads
\begin{align}
H_{\rm JCH} =  \sum^L_{i=1}H^{\rm JC}_i-J\sum_{\langle
  i,j\rangle}A_{ij}a^{\dag}_ia_j+{\rm h.c.}-\sum^{L}_{i=1}\mu_in_i, 
\label{HJCH}
\end{align}
where $L$ is the number of lattice sites, $a_i (a_i^{\dag})$ is the
annihilation (creation) bosonic operator, $J$ is the photon-photon
hopping amplitude, $A_{ij}$ is the adjacency matrix which takes values
$A_{ij}=1$ if two sites of the lattice are connected and $A_{ij}=0$
otherwise. $\mu_i$ stands for the chemical potential at site $i$ and
$n_i=a^{\dag}_ia_i+\sigma^+_i\sigma^-_i$ represents the number of
polaritonic excitations at site $i$. Also,
$H^{\rm JC}_i=\omega a_i^{\dag}a_i +
\omega_0\sigma_i^+\sigma_i^-+g(\sigma_i^+a_i+\sigma_i^-a_i^{\dag})$ is
the Jaynes-Cummings (JC) Hamiltonian describing light-matter
interaction \cite{JC_Model}. Here, $\sigma_i^{+}(\sigma_i^{-})$ is the
raising (lowering) operator acting on the TLS Hilbert space, and
$\omega$, $\omega_0$, and $g$ are the resonator frequency, TLS
frequency, and light-matter coupling strength, respectively. Notice
that the total number of elementary excitations (polaritons) in this
system $N = \sum_i^{M} (a_i^{\dag}a_i + \sigma_i^{\dag}\sigma_i)$ is
the conserved quantity $[N,H_{JCH}]=0$
\cite{hartmann_quantum_2008,hartmann_strong_2007}.

The quantum dynamics of this model has been studied for linear
lattices \cite{Hartmann2006aa,PhysRevA76031805}, and its equilibrium
properties at zero temperature have been studied by means of density
matrix renormalization group \cite{EPLRossini}, and by means of mean
field (MF) theory, for two-dimensional lattices
\cite{NatPhys2006,PhysRevA77031803,PhysRevA80023811} and complex
networks \cite{PhysRevE87022104}. The latter studies have provided
evidence of a quantum phase transition from Mott-insulating phases to
a superfluid polaritonic phase. Beyond the MF approach there have been
important contributions from the numerical and analytical viewpoint
for extracting the phase boundaries
\cite{PhysRevLett99186401,PhysRevLett100216401,arxiv_08063603,%
  PhysRevA80033612,PhysRevLett103086403}, the study of critical
behavior \cite{arxiv_08063603,PhysRevA80033612,PhysRevLett103086403},
and the excitation spectrum
\cite{PhysRevLett100216401,PhysRevA80033612,PhysRevLett103086403}. For
a general overview on many-body physics with light relevant literature
is
available~\cite{Review_Polaritons1,Review_Polaritons2,QSbook_Polaritons}.

\section*{Mott-insulator to superfluid phase transition}
Here we briefly summarize the Mott-insulator to superfluid phase
transition in the JCH model \cite{PhysRevA76031805}. Our main
results are focused on the quantum dynamics of the
JCH model~(\ref{HJCH}) in complex networks, where we focus on 
the canonical ensemble with a fixed total number of polaritons \cite{PhysRevX4031043,PhysRevX7011016}.
In this case, the JCH
Hamiltonian reads
\begin{align}
\centering
H_{\rm JCH} =  \sum^L_{i=1}H^{\rm JC}_i-J\sum_{\langle
  i,j\rangle}A_{ij}a^{\dag}_ia_j +{\rm h.c.}
\label{HJCH1} 
\end{align} 
In the atomic limit, where  the photon-hopping can be neglected
($J\ll g$), the JC Hamiltonian at site $i$ ($H^{\rm JC}_i$) can be
diagonalized in the polaritonic basis that mixes atomic and photonic
excitations
$\ket{n,\pm}_i=\gamma_{n\pm}\ket{\downarrow,n}_i+\rho_{n\pm}\ket{\uparrow,n-1}_i$
with energies $\epsilon^{\pm}_n=n\omega+\Delta/2\pm\chi(n)$, where
$\chi(n)=\sqrt{\Delta^2/4+g^2n}$, $\rho_{n+}=\cos(\theta_n/2)$,
$\gamma_{n+}=\sin(\theta_n/2)$, $\rho_{n-}=-\gamma_{n+}$,
$\gamma_{n-}=\rho_{n+}$, $\tan\theta_n=2g\sqrt{n}/\Delta$, and the
detuning parameter $\Delta=\omega_0-\omega$.

Now, one can introduce the polaritonic creation operators at site $i$
defined as $P^{\dag (n,\alpha)}_i=\ket{n,\alpha}_i\bra{0,-}$, where
$\alpha=\pm$ and we identify $\ket{0,-}\equiv\ket{\downarrow,0}$ and
$\ket{0,+}\equiv\ket{\emptyset}$ being a ket with all entries equal to
zero, that is, it represents an unphysical state. These
identifications imply $\gamma_{0-}=1$ and
$\gamma_{0+}=\rho_{0\pm}=0$. Using this polaritonic mapping the
Hamiltonian (\ref{HJCH1}) can be rewritten as
\cite{PhysRevA76031805,PhysRevA80023811}
\begin{align}
\centering
H =& \sum^L_{i=1}\sum^{\infty}_{n=1}\sum_{\alpha=\pm}\epsilon_n^{\alpha} 
P^{\dag (n,\alpha)}_iP^{(n,\alpha)}_i-J\sum_{\langle i,j\rangle}A_{ij}\Big[\sum^{\infty}_{n,m=1}\sum_{\alpha,\alpha',\beta,\beta'}t^{n}_{\alpha,\alpha'}
t^{m}_{\beta,\beta'}P^{\dag (n-1,\alpha)}_iP^{(n,\alpha')}_iP^{\dag (m,\beta)}_jP^{(m-1,\beta')}_j\nonumber\\
&+ {\rm h.c.}\Big],
\label{HPolariton}
\end{align}
where the matrix elements $t^{n}_{\alpha,\alpha'}$ are given by
$t^{n}_{\pm+}=\sqrt{n}\gamma_{n\pm}\gamma_{(n-1)+}+\sqrt{n-1}\rho_{n\pm}\gamma_{(n-1)-}$
and
$t^{n}_{\pm-}=\sqrt{n}\gamma_{n\pm}\rho_{(n-1)+}+\sqrt{n-1}\rho_{n\pm}\rho_{(n-1)-}$.
The first term in Eq.~(\ref{HPolariton}) stands for the local
polaritonic energy with an anharmonic spectrum and gives rise to an
effective on-site polaritonic repulsion. The last term in
Eq.~(\ref{HPolariton}) represents the polariton hopping between
nearest neighbors and long range sites, and it may also allow for the
interchange of polaritonic excitations.

If the physical parameters of the Hamiltonian (\ref{HPolariton}) are
in the regime $Jn\ll g\sqrt{n}\ll\omega$, and for an integer filling
factor, where the total number of excitations $N$ over the lattice is
an integer multiple of the number of unit cells $L$, the lowest energy
state is the product $\bigotimes_{i=1}^L\ket{1,-}_i$ which corresponds
to a Mott-insulating phase, and its associated energy is
$E=N\epsilon_1^-$. In the thermodynamics limit, the interplay between
the on-site polariton repulsion and the polariton hopping leads to a
phase transition from a Mott insulator to a superfluid phase. The
latter may be reached by diminishing the on-site repulsion by means of
detuning the atomic and photonic frequencies. At equilibrium, this
phase transition may be quantified by means of bipartite fluctuations
\cite{PRLRachel, EPLRossini}. In a simulated Mott-insulator transition, where an
adiabatic dynamics drives the passage, it has been shown that a
suitable order parameter corresponds to the variance of the number of
excitations per site. Fig.~\ref{Fig1}(b) shows the archetypal behavior
of the order parameter as a function of the detuning $\Delta$ in the
adiabatic dynamic regime, and for an integer filling factor of one net
excitation per site \cite{PhysRevA76031805}.

\section*{Quenched dynamics and Topology in finite-size complex
  lattices.} 

Our aim is to describe how complex arrangements of CRAs, such as the
one appearing in Fig.~\ref{Fig1}(a), affect the simulated phase transition from
Mott insulator to superfluid as the detuning parameter $\Delta$ is
suddenly quenched. In particular, we are interested in how one can
manipulate photonic transport properties of the emerging superfluid
phase depending on the specific topology of the CRAs. As order
parameter we choose the time-averaged standard deviation of the
polariton number
$ \frac{1}{T} \int_0^T dt \sum_{i}^L( \langle n_i^2\rangle - \langle
n_i \rangle^2 ))$ with $T=J^{-1}$, and we assume the whole system
initially prepared in the Mott-insulating state
$\ket{\psi_0}=\bigotimes_{i=1}^L\ket{1,-}_i$, with $\Delta=0$ at each
lattice site. In the supplementary material we present another equivalent
measures of the order parameter based on the bipartite fluctuation proposed 
by S. Rachel {\it et al.}\cite{PRLRachel}, and D. Rossini {\it et al.}\cite{EPLRossini}.
Of course, due to computational restrictions, we consider relatively small arrangements 
of CRAs, but with varying degrees of complexity, suggesting that the topology 
of the network could be used in a nontrivial way to manipulate the emerging of 
the superfluid phase as these system becomes larger and approach the thermodynamical
limit.

The initialization process may be achieved by the scheme proposed by
Angelakis et al. \cite{PhysRevA76031805}. For instance, in circuit QED
\cite{PhysRevX4031043,PhysRevX7011016} one might cool down the whole
system reaching temperatures around $T_0\sim 15$mK. In this case, the
system will be prepared in its global ground state
$\ket{G}=\bigotimes_{i=1}^L\ket{0,-}_i$. Then, one can apply
individual magnetic fields on the TLSs, each implemented via a
transmon qubit \cite{PhysRevA76042319}, such that the resonance
condition $\Delta=0$ is achieved.  This way one can address
individually each cavity with an external AC microwave current or
voltage tuned to the transition $\ket{\downarrow,0}_i\to\ket{1,-}_i$,
with a driving frequency $\omega_D=\omega-g$, such that the system
will be prepared in the desired initial state $\ket{\psi_0}.$ The
sudden quench of the detuning can be achieved by applying magnetic
fields to the transmon qubits in order to reach the desired superfluid
phase.  It is noteworthy that when the initial state is a linear
superposition of upper and lower polariton states ($\Delta \neq 0$)
the quantum dynamics will be dominated by these two polaritonic
bands. Also, we carry out full numerical calculations for the
parameters $g=10^{-2}\omega$ and $J=10^{-2}g$, and we consider up to
$6$ Fock states per bosonic mode.  These parameter values allow us to
prevent the interchange of polaritonic excitations between different
sites.
\begin{figure}[t]
\centering
\includegraphics[scale=0.47]{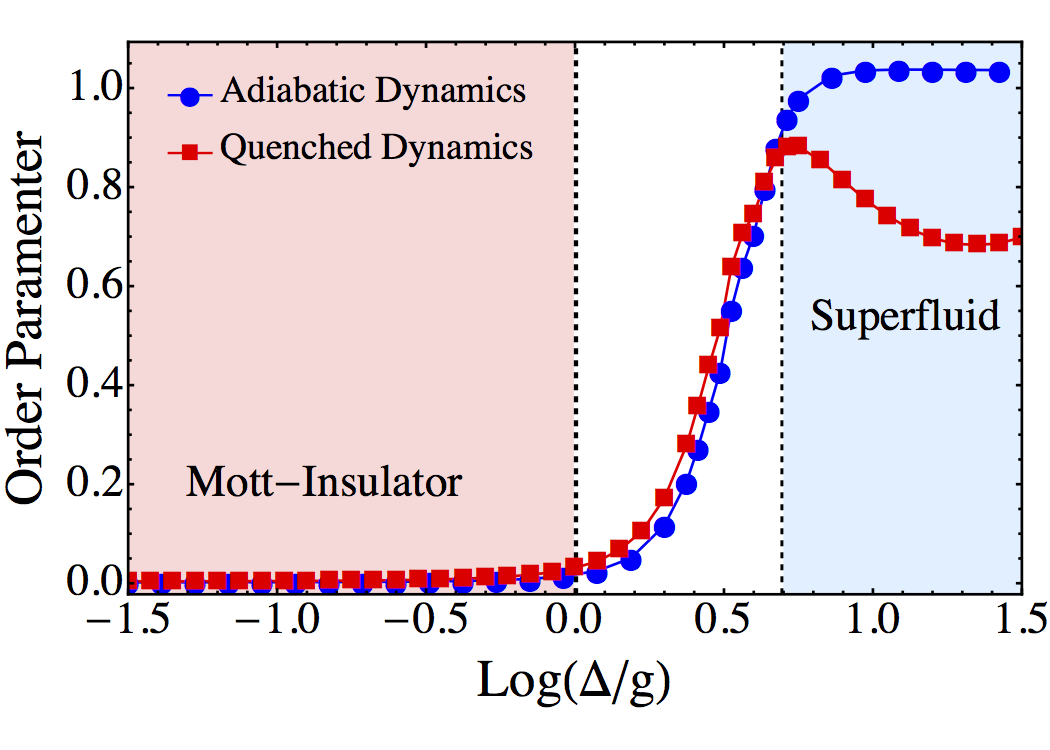}
\caption{(color online) Quantum phase transition of a dimer array.
  Detuning dependence of the order parameter with two TLS coupled
  through photon hopping, adiabatic dynamics (blue circles) and
  quench dynamics (red squares).  Continuous lines have been
  added as guide to the eye.}
\label{Fig2}
\end{figure}

\begin{figure}[htb]
  \centering
\includegraphics[scale=0.4]{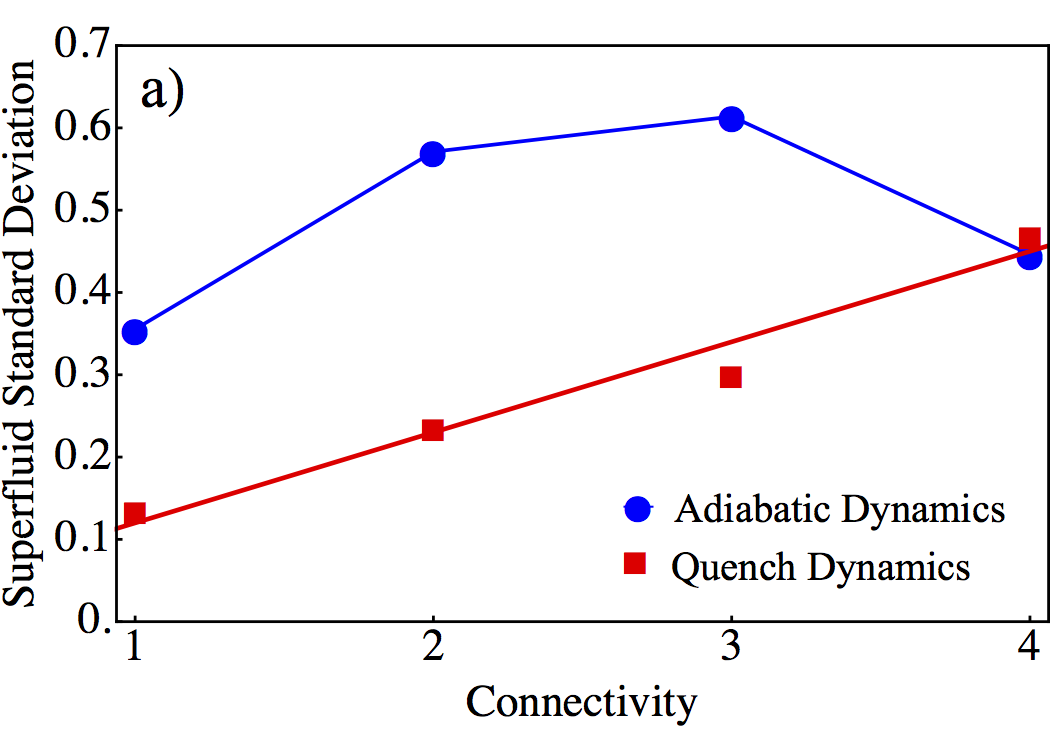}
\includegraphics[scale=0.4]{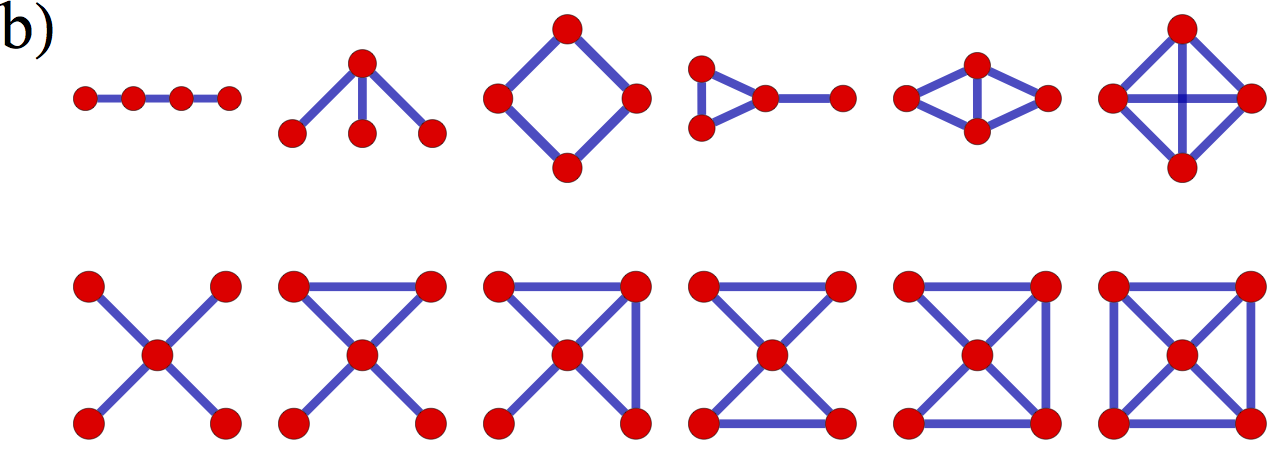}
\caption{(color online) a)~Standard deviation of the superfluid phase
  as a function of the connectivity.  Adiabatic dynamics (blue
  circles) and quench dynamics (red squares). A set of CRAs with four
  and five interconnecting resonators, as shown in b) are
  considered. Continuous lines have been added as a guide to the eye.}
\label{Fig3}
\end{figure}

In order to gain insight into the quench dynamics of the topological
CRAs let us consider a dimer array. As shown in Fig.~\ref{Fig2}, the
simulated Mott-insulator to Superfluid phase transition strongly
depends on the type of dynamics. Adiabatic dynamics resembles a second order
phase transition which leads to a continuous change of the state of
the system. On the other hand, the quench dynamics takes place
accompanied by a discontinuous change of the state, analogous to the
Metal-Insulator transition of oxides \cite{PhysRevB.55.R4855}.  Hence,
as we expected, the adiabatic dynamics is not qualitatively affected
by the distribution of nearest neighbors. However, the topological
properties of the array dominate a first-order like phase transition
driving the quench dynamics (see Fig.~\ref{Fig2}).  As the degree of
inter-connectivity between the resonators grows the distance between
them rapidly diminishes, and thus local correlations become more
important due to quantum interference effects. If scaled up to the
size of the system, due to the increase in the degrees of freedom,
the numerical simulation time grows exponentially. In the next section we
obtain an empirical scaling law to address this issue.  Indeed, we
demonstrate that the photon propagation in the simulated superfluid
phase strongly depends of the connectivity per site
$k_i=\sum_jA_{ij}$. Let us consider a set of arrays with a fixed
number of TLS. As shown in Fig.~\ref{Fig3}(a) in the quench dynamics
case the averaged standard deviation depends linearly on the
connectivity, which means that depending on the connectivity the local
superfluid states are reached with different detuning scales.  We
consider a set of CRAs with four and five interconnecting resonators
as shown in Fig.~\ref{Fig3}(b). In contrast to these results, the
adiabatic dynamics does not exhibit a monotone or linearly growing
behavior, which leads to a sharper phase transition, as illustrated in
Fig.~\ref{Fig2}.

\section*{Mean-field theory of the Superfluid Phase}
\begin{figure}[t]
\centering
\includegraphics[scale=0.2]{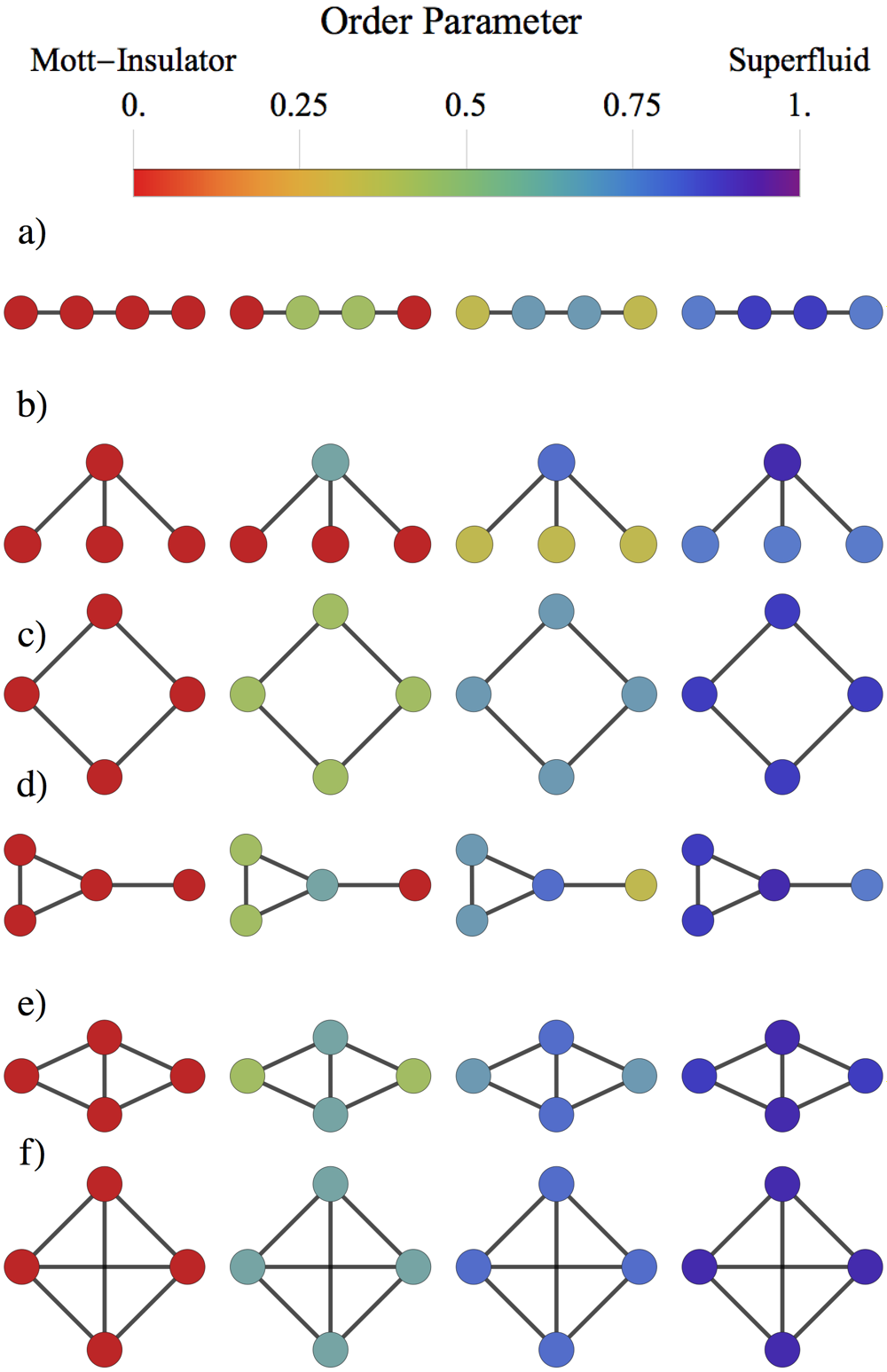}
\caption{(color-online) Numerical simulation of the quench
  dynamics. The full set of four node arrays, with (a-b) three; (c-d)
  four; (e) five; and (f) six connections. Connectivity per site a)
  (1,2,2,1), b) (1,1,1,3), c) (2,2,2,2), d) (1,3,2,2), e) (2,3,3,2),
  and f) (3,3,3,3). As the connectivity is increased locally the
  superfluid phase is achieved with a lower detuning strength.
  For each array and from left to right we have considered parameters
  $\log{(\Delta/g)}=(0.5,0.7,0.75,0.8)$, and $g=10^{-2}\omega$, $J=10^{-3}\omega$, where
  $\omega$ is the resonator frequency.}
\label{Fig4}
\end{figure}

In the thermodynamics limit, the emergent superfluid phase behaves as
a quantum liquid \cite{NatPhys2006}. Superfluidity is achieved by
means of a transition of the excitations from polaritonic to
photonic. In order to describe the simulated superfluid phase in our
system, we introduce the photonic order parameter~\cite{NatPhys2006}
$\psi=\langle a_i \rangle$. Using the decoupling
approximation
$a^{\dagger}_ia_j\approx \langle a^{\dagger}_i \rangle
a_j+a^{\dagger}_i \langle a_j\rangle -\langle a^{\dagger}_i\rangle
\langle a_j\rangle$, the resulting mean-field JCH Hamiltonian can be
written as
\begin{align}
H_{JCH}=\sum_iH^{JC}_i-J\sum_{i}k_i(\psi a^{\dagger}_i+\psi^*a_i).
\end{align}
Therefore, the simulated Mott-insulator phase can be characterized by
the on site repulsion, which suppresses the fluctuations of the number
of per site excitations $|\psi|=0$. On the contrary, the superfluid
phase is dominated by the hopping and the quantum fluctuations
$|\psi|\neq0$.  Now we focus on the light-matter coupling induced by
the hopping of photons through cavities.  Introducing the identity
$\sigma^+\sigma^-+\sigma^-\sigma^+=\mathds{1}$, we obtain an effective
light-matter coupling, since it retains the mixed products of photonic
and two level operators,
\begin{eqnarray}
h^{LM}_i&=&\tilde{g}_ia^{\dagger}_i\sigma^-_i+\tilde{g}^{\dagger}_ia_i\sigma^+_i
            +{\rm h.c.} 
\end{eqnarray}
Here $\tilde{g}_i=\mathds{1} g/2-Jk_i\psi^* \sigma^-_i$ is the effective
light-matter coupling per site, which therefore turns out to be an
operator. In the simulated superfluid phase the atomic transitions are expected
to be suppressed against the photonic dressed states. Moreover, the
total excitation number does not change, hence when the photonic
excitations  increase the atomic excitations decrease.  Note that
when $\tilde{g}_i=\mathds{1} g$, i.e. when there are no hopping or
topological effects,
\begin{equation}
\langle \sigma^+_i\rangle=\dfrac{g}{Jk_i}\dfrac{1}{\psi},
\label{eq:sigma}
\end{equation}
which indicates that the total number of excitations is conserved and
also demonstrates that the increase of the photonic states leads to a
reduction of the atomic excitations, due to the conservation of
the number excitations. Fig.~\ref{Fig4} shows the effect of the quench
dynamics on the simulated phase transition of the JCH model for
different arrays. In this case the nucleation of superfluid states
emerges due to the variation of the order parameter, according to
Eq.~(\ref{eq:sigma}). In the Mott-Insulator state
$\langle \sigma^+_i\rangle>0 \, \forall i$, when the detuning is
increased $\langle \sigma^+_i\rangle$ decreases by a factor
$1/(k_i\psi)$, until the superfluid phase is reached.

We have shown that the mean field approach strongly supports the
scaling law of the order parameter shown in Fig.~\ref{Fig3}(a); namely,
as the connectivity of CRAs is increased locally, the light superfluid phase
is achieved for a smaller detuning strength.

\section*{Conclusion}
We show that quench dynamics induce a first-order like phase
transition in coupled resonator arrays doped with a two-level
system. The nucleation of simulated superfluid states has been
demonstrated by numerical simulation and by a mean field theoretical
approach. In the quench dynamics the abrupt change of the order
parameter, instead of sharper crossover driven by adiabatic dynamics,
is explained by the non uniform transition from Mott-Insulator to
superfluid, which locally depends of the connectivity. Since the
quench dynamics exhibits the same behavior independent of the choice
of the order parameter, the standard deviation of the polariton number
or the bipartite fluctutation, our results reveal the universality of
the simulated first order phase transition (also see Supplementary
Material).  As the number of TLS is increased the averaged standard
deviation of the superfluid phase depends linearly on the
connectivity. At an increased scale, for large networks of doped
optical/microwave resonators, our system may enter the field of
quantum simulators. In particular, as far as we understand, there is
no known microscopic mechanism for predicting nucleation in
first-order phase transitions. In this context, our results provide an
exact geometrical description for the appearance of domain nucleation
due to the number of connections. Thus, our results may be used to predict,
and manipulate, the nucleation of a superfluid phase of light in
complex-random networks.



\section*{Acknowledgements} This work was supported by the Fondo
Nacional de Investigaciones Cient\'ificas y Tecnol\'ogicas (FONDECYT,
Chile) under grants No. $1150806$ (FT), No. 1160639 (MK,JR), $1150718$
(JAV), $1150653$ (GR), Grant-FA9550-16-1-0122 (FT,MK), CEDENNA
through the ``Financiamiento Basal para Centros Cient\'ificos y
Tecnol\'ogicos de Excelencia-FB0807'' and AWARD NO. FA9550-18-1-0438 (FT, JR, MK and JAV).

\section*{Supplementary Material: Nucleation of superfluid-light domains in a quenched dynamics}

\section*{Integration time}

In the definition of the order parameter, namely, using the variance
of the polariton number or the bipartite fluctuations of the polariton
number, we integrate fluctuations up to a time $T=1/J$, which is a
characteristic time scale where the hopping strength dominates the
dynamics, thus producing a delocalized wave function over different
lattice points.  \vspace{0.3cm}

\begin{figure}[h!]
\includegraphics[scale=0.35]{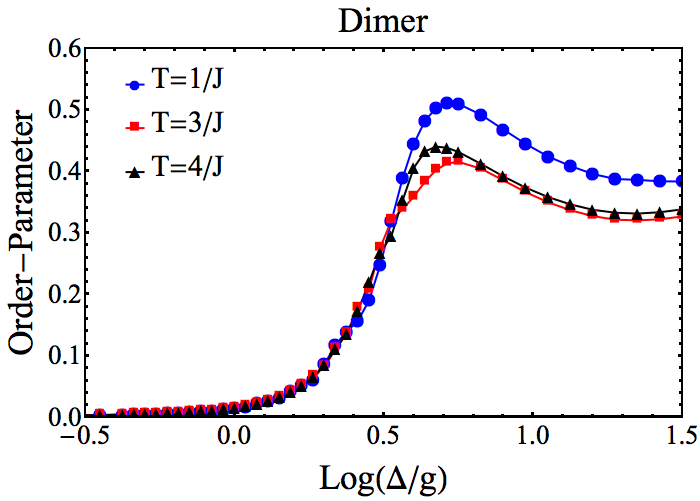}
\includegraphics[scale=0.35]{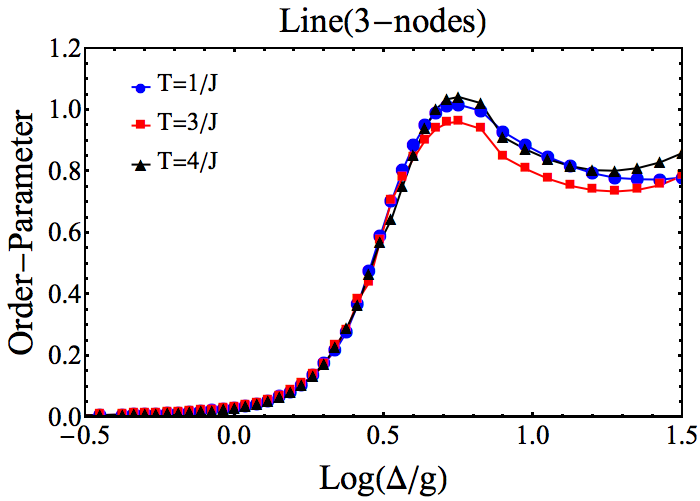}
\caption{(color-online) Order parameter as a function of the detunig
  for different integration times ($T=1/J, 3/J$, and $4/J$). As the
  size of the array is increased there is no significant change in the
  standard deviation of the superfluid phase.}
 \label{fig:time}
\end{figure}

However, for longer simulation times, there are no significant
differences in the behavior of the studied order parameter. We have
carried out numerical simulations for the dimer, and for a linear
array of three coupled resonators; the results are shown in
Fig.~(\ref{fig:time}). We see that the order parameter for different
integration times exhibits the same qualitative behavior. Also, for
times $t<1/J$, not shown here, the studied order parameter almost
vanishes and we do not see the crossover of the Mott-insulator to
superfluid.

\subsection*{Describing Simulated First Order Phase Transition using Bipartite Fluctuations}
\vspace{0.3cm}

The polariton number is conserved in our system since it does not
exchange particles with the outside. Hence for both detuning
$\Delta=0$ and $\Delta\gg g$, where $g$ stands for the light-matter
coupling strength, the corresponding (simulated) Mott and superfluid
states are described by Fock states. This way, if one considers the
standard order parameter studied in mean-field approximation, that is,
the average value of the annihilation operator per site, it will always
be zero and will not capture any crossover from Mott insulator to
superfluid. This is why we choose to study the simulated phase
transition using the onsite variance of the polariton number. However,
we demonstrate that the first-order like-behavior is universal, in the
sense that it does not depend on the choice of the order
parameter. Indeed, if we use the bipartite fluctuations of subsystems
we find that this approach also provides a correct description of the Mott to
superfluid phase transition. Certainly, the dispersion of the polariton number
on a given partition $M-th$ can be assessed by the variance of this 
subsystem, which can be obtained from the two-point correlation 
function $C_{i,j}=\langle n_in_j\rangle-\langle n_i\rangle\langle n_j\rangle$,
using the parameter $\sum_{i,j\in M}C_{ij}$, where $n_i$ denotes the polariton 
occupation number at the i$-th$ site, \cite{PRLRachel, EPLRossini}.

\vspace{0.3cm}

\begin{figure}[h!]
\includegraphics[scale=0.33]{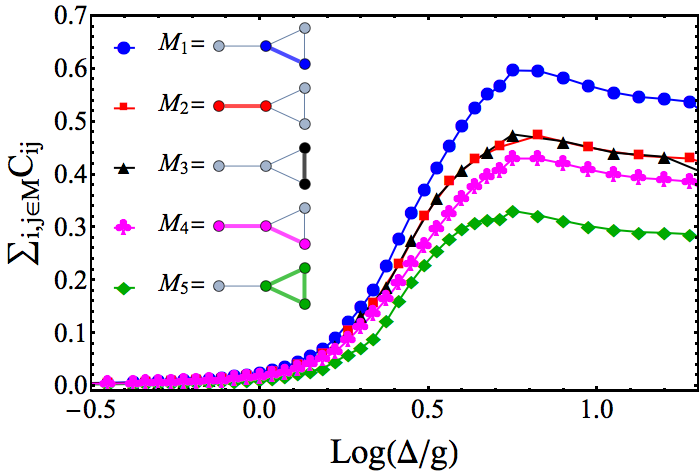}
\includegraphics[scale=0.33]{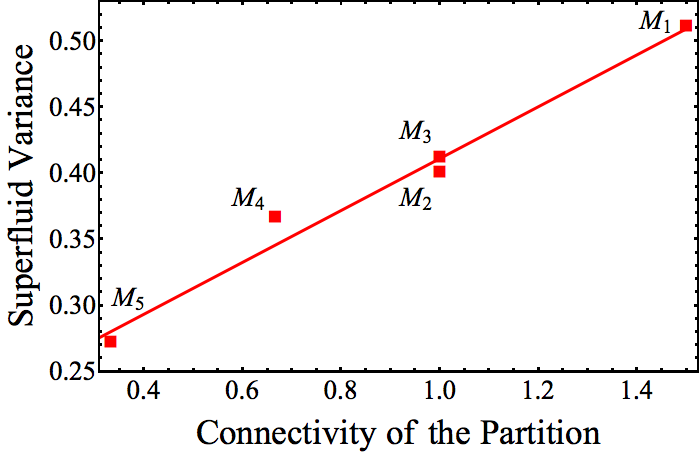}
\caption{(color-online) (left) Bipartite fluctuations of polariton
  occupation number belonging to the M-th partition as a function of
  detunning. (right) Standard deviation of the superfluid phase as
  a function of the connectivity of the M-th partition.}
 \label{fig:BH}
\end{figure}

We have performed numerical calculations for bipartite fluctuations in
our finite arrays. Thus we can demonstrate that a simulated first
order phase transition appears regardless of the choice of the order
parameter; see left-panel of Fig.~(\ref{fig:BH}).  As shown in the
right-panel of the Fig.~(\ref{fig:BH}), using the bipartite fluctuations
approach, the averaged standard deviation also displays a linear
dependence on the connectivity of the partition. 

Where the
connectivity of the partition has been defined as the ratio between
the external connectivity (total number of links between one node in
the partition and the external nodes) of each node and the total
number of nodes. The connectivity of the partition are summarized 
in Table 1. These results demonstrated the linear dependence of 
the order between the connectivity and the order parameter, in both 
case, when the order parameter is identified as the standard deviation 
or as the variance of the bipartite fluctuations.

\begin{table}[ht]
\centering
	\begin{tabular}{| c | c | c | c |}
	\hline
	Partition & External connectivity & Number of Nodes & Connectivity of the partition\\
	\hline
	$M_1$ & 2+1 & 2 & $3/2$ \\
	\hline
	$M_2$ & 2 & 2 & $2/2$ \\
	\hline
	$M_3$ & 2 & 2 &  $2/2$ \\
	\hline
	$M_4$ & 2 & 3 &  $2/3$ \\
	\hline
	$M_5$ & 1 & 3 & $1/3$ \\
	\hline
	\end{tabular}
\caption{Summary of the connectivity of each partition corresponding to the array of the 
inset of Fig.~(\ref{fig:BH}-left). The variance of the superfluid phase depends linearly 
on the connectivity of the partition, as shown in Fig.~(\ref{fig:BH}-right)  }
\end{table}

\end{document}